\documentclass[aps, prl, twocolumn]{revtex4-2}
\usepackage{graphicx,bm,float,amsthm,amsmath}
\usepackage{color, colortbl,ulem,mdsymbol,tikz,adjustbox}
\usepackage[utf8]{inputenc}
\usetikzlibrary{decorations.pathreplacing,calligraphy}
\tikzset{every node/.style={align=center}} 
\newcommand\numberthis{\addtocounter{equation}{1}\tag{\theequation}}

\begin{document}
\title{Experimenting quantum phenomena on NISQ computers using high level quantum programming}

\author{Duc M. Tran}
\affiliation{Nano and Energy Center, VNU University of Science, Vietnam National University, 120401 Hanoi, Vietnam}
\author{Duy V. Nguyen}
\affiliation{Phenikaa Institute for Advanced Study, Phenikaa University, 12116 Hanoi, Vietnam}
\affiliation{Faculty of Computer Science, Phenikaa University, 12116 Hanoi, Vietnam}
\author{Le Bin Ho}
\affiliation{Center of Physics, Institute of Applied Mechanics and Informatics, Vietnam Academy of Science and Technology, Ho Chi Minh City 70000, Vietnam}
\affiliation{Research Institute of Electrical Communication, Tohoku University, Sendai 980-8577, Japan}
\author{Hung Q. Nguyen}
\email{hungngq@hus.edu.vn}
\affiliation{Nano and Energy Center, VNU University of Science, Vietnam National University, 120401 Hanoi, Vietnam}

\begin{abstract}
  We execute the quantum eraser, the Elitzur-Vaidman bomb, and the Hardy's paradox experiment using high-level programming language on a generic, gate-based superconducting quantum processor made publicly available by IBM. The quantum circuits for these experiments use a mixture of one-qubit and multi-qubit gates and require high entanglement gate accuracy. The results aligned with theoretical predictions of quantum mechanics to high confidence on circuits using up to 3 qubits. The power of quantum computers and high-level language as a platform for experimenting and studying quantum phenomena is henceforth demonstrated.
\end{abstract}
\maketitle

\section*{Introduction}
Recent advances in quantum engineering have brought about the realization of quantum computers that perform key quantum algorithms satisfactorily and display gigantic potentials \cite{GoogleSupremacy,ChinaSupremacy}. This powerful Turing machine employs quantum superposition and entanglement to perform calculations. By entangling multiple qubits together, techniques such as quantum parallelism is achieved to provide exponential speedup for quantum algorithms over classical ones. Despite prospective impacts, it is tremendously difficult to build a practical quantum computer that is fault-tolerant \cite{DiVincenzo}. The main technical obstacle lies in the fact that a quantum state is fragile to noise. It is hard to engineer a quantum system that is both well-controlled and long-lived. Termed NISQ for Noisy Intermediate Scale Quantum \cite{PreskillNISQ}, these prototypes are not ready for real-life applications that require complexity and precision lying beyond our current state of the art.

At the moment, the leading candidate in hardware is perhaps superconducting circuits \cite{Kjaergaard}. At its heart, the qubit comprises a collective state of electrons, condensed into a single quantum state, the Cooper pairs. This artificial quantum object can be freely designed and engineered, conveniently controlled using microwave pulses, and scalable using established nano-fabrications techniques. In this system, for example of a charge qubit \cite{Nakamura}, the quantum superposition is realized by a microwave pulse that places the tiny superconducting island at a potential that supports both $N$ and $N+1$ number of charges at the same time. This action is equivalent to a beamsplitter in a Mach-Zehnder interferometer (MZI) that splits the photon's route into two uncertain paths. In the quantum circuit language, this is performed by a Hadamard gate with matrix form $H=\frac{1}{\sqrt{2}}\bigl[\begin{smallmatrix} 1 & 1 \\ 1 & -1 \end{smallmatrix}\bigr]$. In a transmon quantum processor \cite{transmon}, such as those provided by IBM \cite{IBMQ}, the error of a Hadamard gate is in the range of $10^{-3}$ to $10^{-4}$. Designed as a universal computing platform and programmed using high-level language through cloud access, there is a range of remarkable works performed on real devices with few qubits. They include the demonstration of important quantum algorithms \cite{ShorIBM,GroverIBM}, simulation of quantum phenomena \cite{SolanoIon,MartinisNComm,Neutrino,BlattNature,HouckQED,Gambetta17}, or reproducing foundation quantum experiences \cite{Sierra5,IBMBell,IBMduality,IBMcat,DevittPRA}.

Quantum physics is a generic theory that applies to any physical objects, such as photons, electrons, or artificial devices engineered in the lab. However, lying on the surface of a chip, the superconducting qubits have limited connectivity between neighbors. It is unclear if the entanglement of a collective quantum state induced by a metallic coupler is the same as that of genuine nature elements. In this work, we perform three fundamental experiments - those at the core of quantum physics - by programming the IBM NISQ computers. Using Python scripts written in Qiskit, we demonstrate that the quantum eraser, the Elitzur-Vaidman bomb, and the Hardy's paradox experiments can be reproduced successfully on the same quantum processor. Originally proposed and realized in optical apparatuses, we showed that theoretically, these optical setups are equivalent to our quantum circuits. The quantum circuits are executed on real quantum computers with reasonable agreement \cite{github}. The experimental deviations to the original results are associated with the erroneousness in the performance of the superconducting devices provided by IBM.

\section*{The quantum eraser}
\begin{figure}[t]
\begin{center}
\includegraphics[width=3.3in,keepaspectratio]{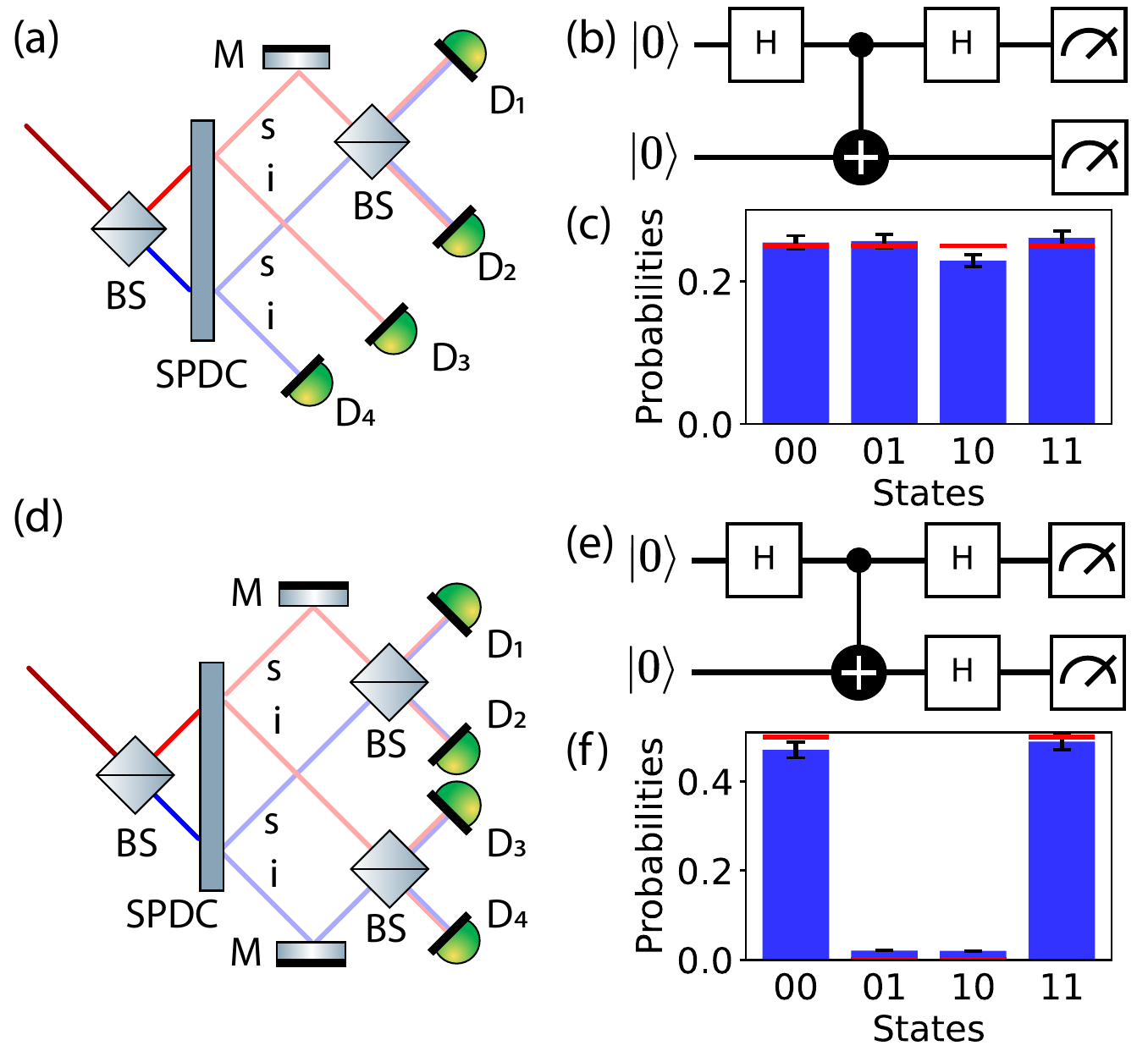}
    \caption{In the quantum eraser experiment, we recover the interference pattern by choosing an appropriate measurement basis. (a, d) The quantum eraser experiments using a Mach-Zehnder interferometer (MZI) and a spontaneous parametric down-conversion (SPDC), an alternative setup to the double-slit experiment: upon entering the first beamsplitter (BS), the original photon is split into upper-path (red) and lower-path (blue). Each path transforms into a signal-idler photons pair by passing through the SPDC. The signal photons (s) interfere at the second BS and are measured by D$_{1(2)}$. The idler photons are measured by D$_{3(4)}$ without an eraser in (a) and with an eraser in (d).  (b, e) Quantum circuit for the experiment in (a, d), respectively. (c, f) Experiment results obtained from the IBM Vigo quantum computer shown as blue bars and the theoretical value showed as red lines. The device error is estimated at 3.7\%, while standard error from the mean of 8192 shots is minimal at 0.5\%.}
  \label{fig1}
\end{center}
\end{figure}

The quantum eraser experiment \cite{Eraser,Scully82,Kim2000} is a natural extension of the double-slit experiment that demonstrates the counter-intuitive nature of quantum physics - the wave-particle duality. In Fig.~\ref{fig1} (a, d), the duality is eavesdropped on by using the Mach-Zehnder interferometer (MZI) apparatus with a spontaneous parametric down-conversion (SPDC) process. Upon passing the first beamsplitter (BS), the photon splits into an upper-path (red) and a lower-path (blue), which indicates the ``which-path" information. These paths encounter the SPDC that generates an entangled ``signal-idler" photons pair for each path, which are denoted as (s) and (i) in Fig.~\ref{fig1} (a, d), respectively. The signals arrive at the second BS to interfere and are detected by D$_1$, D$_2$ detectors. The idlers are measured by D$_3$ and D$_4$ detectors. In Fig.~\ref{fig1} (a), we measure them directly after the SPDC, which reveals the red-path or blue-path information by the clicking of D$_3$ or D$_4$, respectively. In this case, both D$_1$ and D$_2$ click with equal probability, which implies that there is a loss of interference in the signal photons. By observing the which-path information via idler photons, the interference in the signal photons is deleted; in other words, they behave as ``particles." In Fig.~\ref{fig1} (d), the idlers are interfered with a BS before the measurement. Hence, the which-path information is ``erased." Jointly measuring D$_{1,2}$ and D$_{3,4}$ yields surprising results: only D$_1$ and D$_3$ click coincidentally but not for D$_2$ and D$_3$. It indicates that the interference in the signal photons produces a constructive signal at D$_1$ and a destructive signal at D$_2$. Inversely, only D$_2$ and D$_4$ click coincidentally but not for D$_1$ and D$_4$ detectors. The interference in the signal photons generates a constructive signal at D$_2$ and a destructive signal at D$_1$. Thus, by erasing the which-path information, we observe interference in signal photons; in other words, they behave as ``waves." The wave-particle duality depends on ``our action" on the entangled counterparts. Termed quantum eraser, the phenomenon is verified for photons \cite{Kim2000,Kaiser,SlitEraser}, or surface acoustic phonons \cite{Cleland}.

This experiment can be written in the quantum circuit language \cite{IonicioiuPRL} as showed in  Fig.~\ref{fig1} (b, e). Similar to the role of a BS in an optical setup, the first Hadamard gate splits the first qubit's initial state $|q_0\rangle=|0\rangle$ into $|0\rangle$ and $|1\rangle$ to represent the which-path information: $|0\rangle$ $\leftrightarrow$ red-path, and $|1\rangle$ $\leftrightarrow$ blue-path. Then, a CNOT gate entangles $|q_0\rangle$ with the target $|q_1\rangle$ qubit, which simulates the function of the SPDC. By resembling $|q_0\rangle$ as the signals, a second Hadamard gate is applied on it to probe the interference patterns. The outcome of this quantum eraser experiment relies on qubit $|q_1\rangle$, which plays the role of the idler photons. In Fig.~\ref{fig1} (b), $|q_1\rangle$ is measured right after the CNOT gate such that the which-path information can be verified. The measurement results are shown in Fig.~\ref{fig1} (c), indicating no interference in both $|q_0\rangle$ and $|q_1\rangle$. Similarly, in Fig.~\ref{fig1} (e), $|q_1\rangle$ is measured after the action of the Hadamard gate. This is equivalent to the application of a BS on idler photons in Fig.~\ref{fig1} (d). The which-path information is erased and interference patterns are detected in $|q_0\rangle$, as can be seen from Fig.~\ref{fig1} (f). Mathematically, it is simple to verify the equivalence between the quantum circuit and the optical MZI setup. Using matrix multiplication, it is straightforward to confirm that the circuit in Fig.~\ref{fig1} (b) changes state $|00\rangle$ to $\frac{1}{2}(|00\rangle+|01\rangle+|10\rangle-|11\rangle$, while the circuit in Fig.~\ref{fig1} (e) changes $|00\rangle$ to $\frac{1}{\sqrt{2}}(|00\rangle+|11\rangle$, in agreement with the optical MZI setup in Fig.~\ref{fig1} (a, d), respectively. These results are confirmed by simulation results as solid lines on Fig.~\ref{fig1} (c, f) and agree with the original proposal in the optical apparatus. We execute these circuits on IBM's Vigo quantum computer by programming them using Qiskit \cite{github}. In the non-erased experiment Fig. 1c, the measurement probabilities for $|00\rangle$, $|01\rangle$, $|10\rangle$, and $|11\rangle$ are $25.4\pm 0.9\%, 25.6\pm 0.9\%, 22.9\pm 0.8\%, 26.1\pm 1.0\%$, respectively. Compared to theoretical value of 25\% for each state, the experiment results are quite closely matched. Similarly in the erased experiment Fig. 1f, the corresponding probabilities are $47.0\pm 1.7\%, 2.1\pm 0.1\%, 2.0\pm 0.1\%, 48.9\pm 1.8\%$, which resemble the  theoretical  distribution of 50\%, 0\%, 0\%, 50\%. It is possible to estimate errors from our quantum circuits in Fig. 1b,e due to gate and measurement errors. Using calibration values provided by IBM \cite{github}, the uncertainty in our circuits is 3.7\%. We emphasize that a classical interpretation would predict no recovery of the interference pattern in the erased case, and an equal distribution to all states in the non-erased case; this was not found. Our result, therefore, satisfactorily demonstrate the quantum nature of the IBM machines.

In the quantum circuit representation, the difference between (b) and (e) is the Hadamard gate on $|q_1\rangle$, which changes the measurement basis: in (b) the measurement is performed on the $z$-basis, while in (e) it is on the $x$-basis. The wave-particle duality is thus demonstrated by the choice of measurement basis. This is a feature of our proposed scheme in contrast with \cite{IonicioiuPRL}, where they encoded the wave and particle behaviors into one photon (qubit) while entangling with another ancilla photon. 

\section*{The Elitzur-Vaidman bomb}

\begin{table*}
	\begin{center}
	\begin{adjustbox}{width=1.5\columnwidth}
	\begin{tabular}{|c|c|c|c|c|c|c|c|}\hline 
		& Essex & Ourense & Burlington & London & Vigo & Valencia & x2  \\
		\hline\hline
		$\eta $ & 0.417 & 0.387 & 0.303 & 0.306 & 0.356 & 0.325 & 0.309 \\
		Absolute error & 0.084 & 0.054 & 0.031 & 0.027 & 0.022 & 0.008 & 0.024 \\
		Relative error & 25.1\% & 16.2\% & 9.2\% & 8.1\% & 6.7\% & 2.5\% & 7.3\% \\
		Standard deviation & 0.034 & 0.047 & 0.060 & 0.029 & 0.017 & 0.025 & 0.028 \\
		Relative device error & 9.3\% & 4.9\% & 6.2\% & 7.4\% & 4.1\% & 4.0\% & 5.3\% \\
		Absolute device error & 0.039 & 0.019 & 0.019 & 0.023 & 0.015 & 0.013 & 0.016 \\
		\hline
	\end{tabular}
	\end{adjustbox}
	\end{center}
	\caption{Performance of IBM's quantum devices in our Elitzur-Vaidman bomb experiments. The averaged efficiency $\eta$ for all machines are listed in the first row, differ from the theoretical efficiency $\eta=1/3$. The absolute error is calculated as $|\eta-1/3|$, relative error is $\frac{|1/3-\eta|}{1/3}$, and standard deviation is $\sqrt{ \sum_{i=1}^n\frac{(\eta_i-\eta)^2}{n}}$, with $n$ is the total number of runs and $\eta_i$ is the efficiency of the $i^{\text{th}}$ run. Device error is calculated using the fidelity of a circuit, which is estimated as the product of fidelities of the circuit's component gates. This  information from IBM can be found in our Github \cite{github} for the time that we ran these experiments.}
	\label{tab:error}
\end{table*}

\begin{figure}[h!]
  \begin{center}
    \includegraphics[width=3.5in,keepaspectratio]{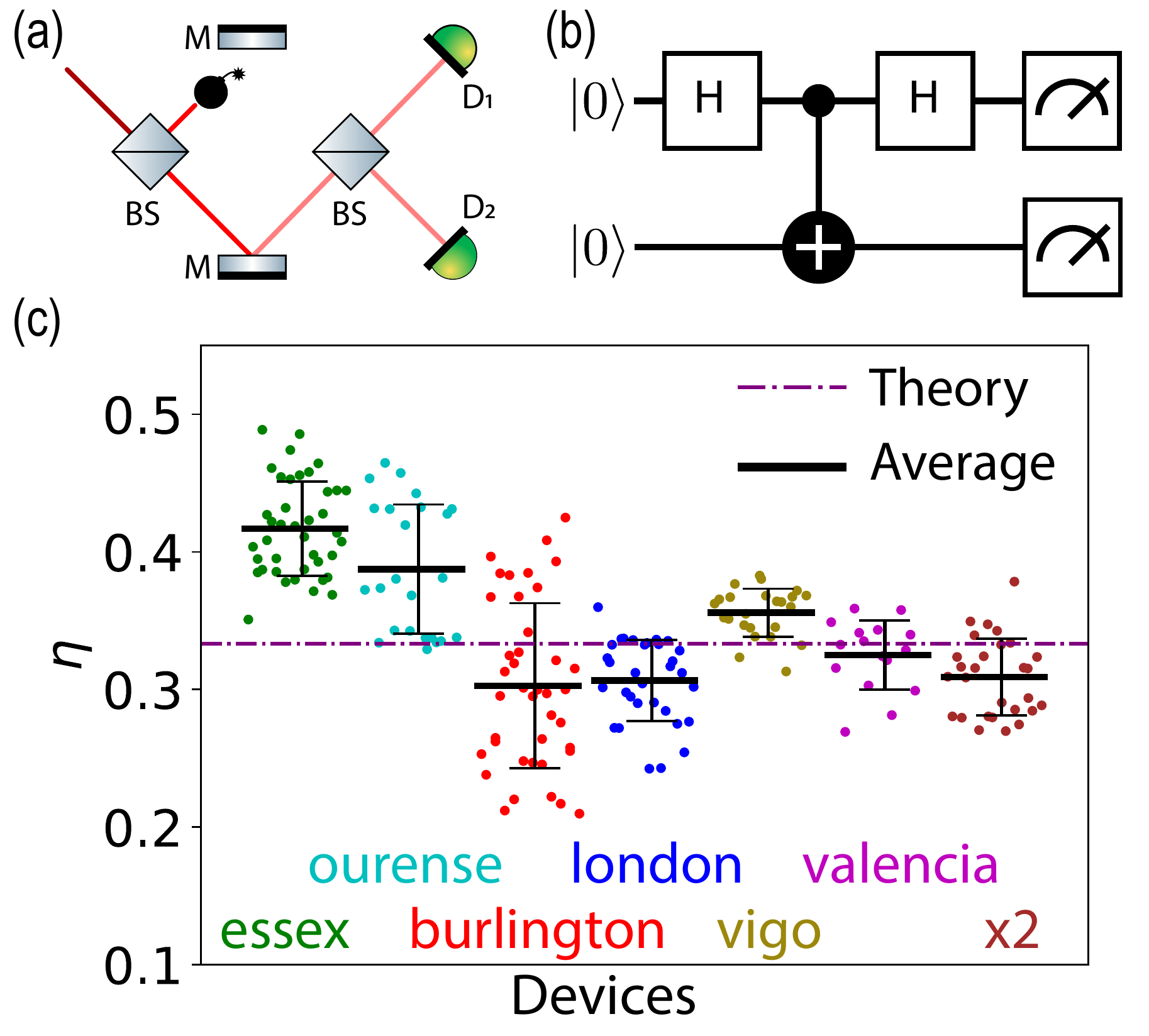}
    \caption{The Elitzur-Vaidman bomb experiment. (a) The original optical setup \cite{EVBomb} with a light-sensitive bomb inside a MZI. There is a non-zero probability that a single measurement detects the bomb without an interaction, i.e., the photon takes the lower path and hits detector D$_1$. More detail description of this apparatus can be found in the text. (b) Quantum circuit for this experiment, where the first qubit represents the photon and the second qubit reveals the status of the bomb. (c) Our experiments on a range of IBM quantum computers in 2020. The dashed-dotted line at 0.33 is the theoretical value from Eq. 1. Each dot represents an execution of the experiment on the corresponding quantum computer. The solid black lines are the average value of dots. The device errors and standard deviations are 9.3\%, 4.9\%, 6.2\%, 7.4\%, 4.1\%, 4.0\%, 5.3\% and 0.034, 0.047, 0.060, 0.029, 0.017, 0.025, 0.028, in order given above, from left to right.}
    \label{fig2}
  \end{center}
\end{figure}

The second experiment we address here is the interaction-free measurement, also known as the Elitzur-Vaidman bomb \cite{EVBomb,Vaidman}. In an optical experiment as shown in Fig.~\ref{fig2} (a), a classical object, such as a light-sensitive bomb, is placed on one path of the MZI. Due to the non-locality of the photon's wavefunction, there is a probability that the bomb is detected without actually being hit by the photon, thus the name interaction-free measurement. This experiment has been verified for photon \cite{Kwiat}, electrons \cite{electronBomb}, atoms \cite{AlbertiBomb}, as well as proposed for solid-state systems \cite{Burkard10,ParaoanuPRL}.

Putting philosophical discussion aside, this experiment can be described as follows. In a standard MZI without the bomb, the original photon splits into an upper path and a lower path with 50\% chance for each case after the first BS. It then interferes with itself at the second BS, and always lights up the constructive detector D$_2$. With the bomb however, if the photon takes the upper path, it will be absorbed by the bomb, and thus produces no clicks in the detectors. If the lower path is taken, the photon will split again with an even chance for each possibility at the second BS, resulting in clicks at D$_1$ and D$_2$ with 
25\% detection each. While the click on D$_2$ is indistinguishable from the no-bomb case, a click on detector D$_1$ tells us that there is an active bomb residing on the upper path inside the MZI. In the latter case, we detect the existence of the bomb without touching it, hence the name ``interaction-free" measurement.

This experiment is described using a quantum circuit containing two Hadamard gates together with a CNOT gate. In our setup as shown in Fig.~\ref{fig2} (b), the first qubit represents the photon in the MZI, and the two Hadamard gates simulate the two BSs. Like the bomb that detects the photon's quantum state inside the MZI, a CNOT gate between the two H gates is targeted on a second qubit. The value of this second qubit $|q_1\rangle$ indicates if the bomb has been touched or not. Let $|q_1\rangle=|1\rangle$ means the bomb is touched, we summarize what happens in the circuit via a diagram as follows:

\begin{adjustbox}{width=\linewidth}
\begin{tikzpicture}[ultra thick,x=1cm,y=0.8cm]
    \node at (0.75, 2) (h1) {$H$};
    \node at (2.5, 2) (cnot) {$CNOT$};
    \node at (4.25, 2) (h2) {$H$};
    \node at (0,0) (a) {$|00\rangle$};
    \node at (1.5, 1) (b) {$|10\rangle$};
    \node at (1.5, -1) (c) {$|00\rangle$};
    \node at (3.5, 1) (d) {$|11\rangle$};
    \node at (3.5, -1) (e) {$|00\rangle$};
    \node at (5, 1.5) (f) {$|01\rangle$};
    \node at (5, 0.5) (g) {$|11\rangle$};
    \node at (5, -0.5) (h) {$|00\rangle$};
    \node at (5, -1.5) (i) {$|10\rangle$};
    \node at (7, 1.25) {touched and};
    \node at (7.25, 0.75) {exploded bomb};
    \node at (7.5, -0.225) {no bomb or cannot};
    \node at (7.35, -0.65) {detect any bomb};
    \node at (7.35, -1.5) {detected a bomb};
    \node at (10, 1) {(50\%)};
    \node at (10, -0.5) {(25\%)};
    \node at (10, -1.5) {(25\%)};
    \draw [->] (a)--(b);
    \draw [->] (a)--(c);
    \draw [->] (b)--(d);
    \draw [->] (c)--(e);
    \draw [->] (d)--(f);
    \draw [->] (d)--(g);
    \draw [->] (e)--(h);
    \draw [->] (e)--(i);
    \draw [decorate, decoration = {calligraphic brace}] (5.5,1.75) --  (5.55,0.25); 
\end{tikzpicture}
\end{adjustbox}

Even though there is a $50\%$ probability of explosion, this scheme allows us to probe the existence of a classical object in an interaction-free manner. The efficiency rate $\eta$ can be expressed by a measurement on the two-qubit system \cite{Kwiat}
\begin{equation}\label{eq:eta}
\eta=\frac{P_{\text{det}}}{P_{\text{det}}+P_{\text{abs}}}=\frac{P_{|10\rangle}}{P_{|10\rangle}+P_{|01\rangle}+P_{|11\rangle}}=\frac{P_{|10\rangle}}{1-P_{|00\rangle}}.
\end{equation}
Here, $P_{\text{det}}$ is the probability of detecting the bomb without touching it, and $P_{\text{abs}}$ is the probability that the bomb explodes; or in other words, the photon is absorbed. With a 50:50 beamsplitter, the efficiency rate $\eta = \frac{1/4}{1/4+1/2} = \frac13$. We execute the quantum circuit using a virtual quantum machine and obtain an exact value of 1/3, shown as a dashed-dotted line in Fig.~\ref{fig2} (c). The same circuit running on real IBMQ quantum hardware at various times in August 2020 yields results that differ from each other and deviate from the theoretical value 1/3. We learn that even when mitigation error correction \cite{mitigation} is applied, most systems have large errors as shown with error bars in Fig. 2c, see Table 1 for more detail for Essex, Ourense, Burlington, London, Vigo, Valencia, and x2.  Especially, IBM Essex is systematically off to one side, while IBM London is off to the other side of the average value. Of all devices, the IBM Vigo and IBM Valencia achieve the best result with the smallest errors in all aspects. We emphasize that these results are obtained from an identical set of codes, and thus imply large variations in IBM's hardware. Even with a shallow circuit as shown in Fig. 2b, one need to be cautious and choose a proper backend for their calculations.

\begin{figure}[h!]
  \begin{center}
    \includegraphics[width=3.4in,keepaspectratio]{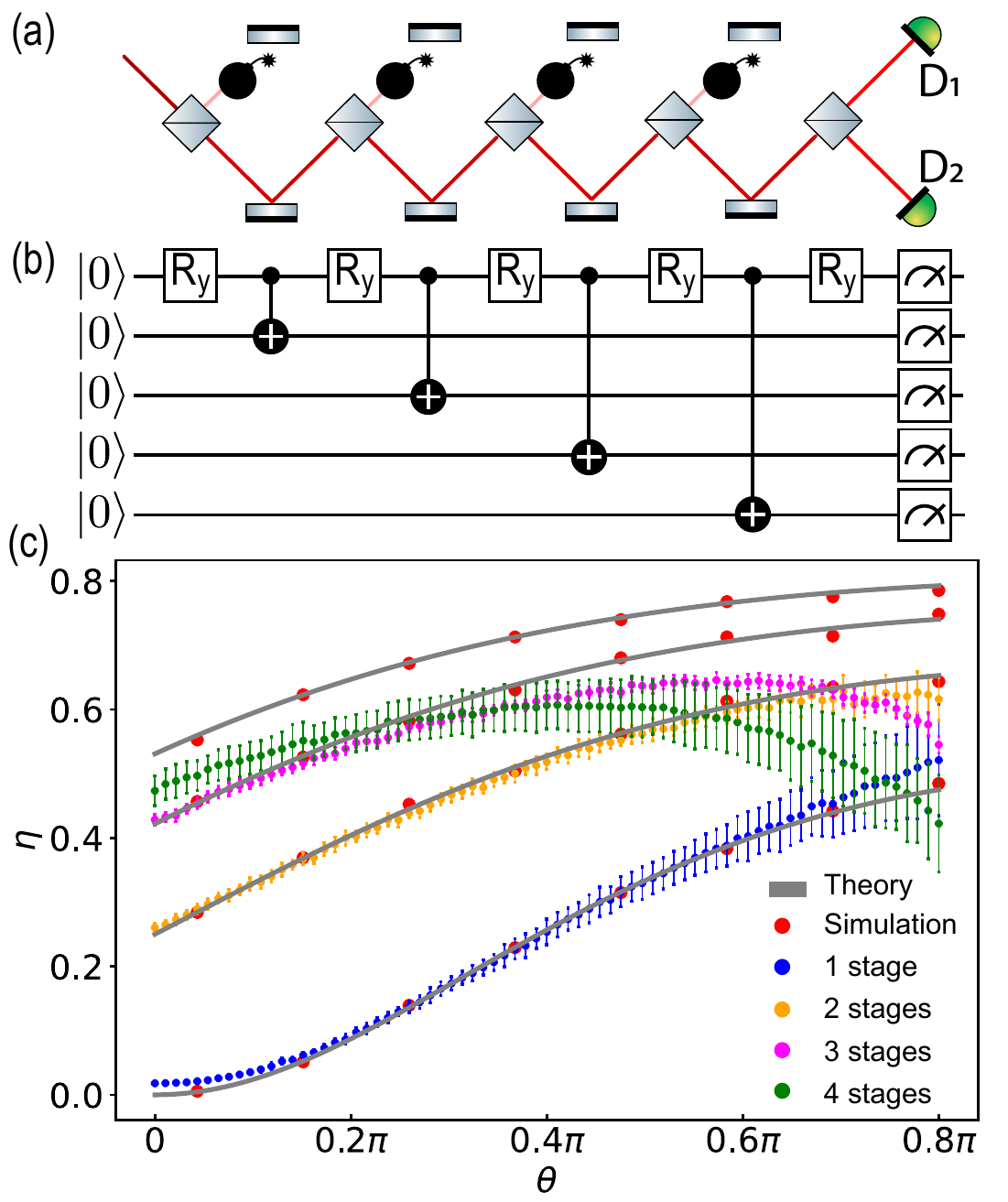}
    \caption{The general Elitzur-Vaidman bomb experiment. (a) The original optical setup \cite{Kwiat} used a series of MZIs to detect the bomb multiple times. (b) Quantum circuit for the experiment. The $R_y(\theta_i)$ replaces the $H$ gate for control over the reflectivity; an extension from Fig. 2b. (c) Our experiment data from the IBMQ Valencia. Since Valencia is a 5-qubit computer, the experiment extends to 4 stages. Device errors are 7.2\%, 10.3\%, 13.8\%, 19.0\% for stage $n=2,3,4,5$, respectively. Standard deviations for all points are shown as error bars, with the largest at 0.087.}
    \label{fig3}
  \end{center}
\end{figure}

To qualify for ``interaction-free measurement"\cite{Kwiat}, the efficiency $\eta$ needs to reach 100\%. In the optical setup, this efficiency can be obtained by a series of connected MZIs such that the photon repeatedly detects the bomb multiple times. In a setup with $N$ BSs, each with  reflectivity $R=\cos^2(\pi/2N)$, the photon gradually transfers from the lower to the upper halves, 
as shown in Fig.~\ref{fig3} (a). After every beamsplitter, there is a dominant chance that the photon takes the lower path and avoids the bomb, and every time it does so the quantum state collapses and fully recovers at the lower path before continuing to evolve unitarily \cite{Vaidman,Misra}. After $N-1$ stages with $N$ beamsplitters, the probability that the photon only takes the lower path becomes $[\cos^2(\pi/2N)]^N$. Following Eq.~\eqref{eq:eta}, the efficiency $\eta$ \cite{Kwiat} is 
\begin{equation}\label{eq:eta2}
  \eta=\frac{\cos^{2N}(\frac{\pi}{2N})}{1-\sin^2(\frac{\pi}{2N})\cos^{2(N-1)}(\frac{\pi}{2N})}.
\end{equation}

The quantum circuit that simulates the series of MZIs in Fig.~\ref{fig3} (a) is shown in Fig.~\ref{fig3} (b) for the case $N = 5$, 
which is a natural expansion from the circuit shown in Fig.~\ref{fig2} (b). The first qubit represents the chain of MZI and the photon that pass through it. The Hadamard gate is now replaced by rotation gates in the $y$-axis $R_y(\theta) = \bigl[\begin{smallmatrix} \cos(\theta/2) & -\sin(\theta/2) \\ \sin(\theta/2) & \cos(\theta/2) \end{smallmatrix}\bigr]$. We assume an arbitrary angle for these rotations $R_{y,i}(\theta_i)$, with $i \in [1,N]$. These angles need to satisfy the condition 
\begin{equation}\label{eq:cond}
  \sum_{i=1}^N\theta_i=\pi,
\end{equation}
such that the measurement is performed on the original basis after all these rotations. In contrast to Hadamard gates, a chain of $R_y(\theta_i)$ satisfying Eq. \eqref{eq:cond} interferes $|0\rangle \rightarrow |1\rangle$, instead of $|0\rangle \rightarrow |0\rangle$. To detect the status of the photon, a chain of CNOT gates is added in between these rotation $R_y$. If any of these target qubits change their value to $|1\rangle$, the photon is absorbed and the bomb explodes. The state of interested is $|00..00\rangle$; it indicates that the photon has successfully passed through the circuit without exploding the bomb. Here, we post-select out the explosion cases, i.e., the target qubits turn to $|1\rangle$. Since the CNOTs do not change the value of their control qubit $q_0$, the coefficient for state $|00..00\rangle$ and $|10..00\rangle$ are $\prod_{i=1}^{N}\cos(\theta_i/2)$ and $\sin(\theta_N/2)\prod_{i=1}^{N-1}\cos(\theta_i/2)$, respectively. From Eq.~\eqref{eq:eta}, $\eta=\frac{P_{\text{det}}}{P_{\text{det}}+P_{\text{abs}}}$, the efficiency becomes
\begin{equation}\label{eq:eta4}
  \eta=\dfrac{P_{|00..00\rangle}}{1-P_{|10..00\rangle}}
  =\frac{\prod_{i=1}^{N}\cos^2\Big(\dfrac{\theta_i}{2}\Big)}{1-\sin^2\Big(\dfrac{\theta_N}{2}\Big)\prod_{i=1}^{N-1}\cos^2\Big(\dfrac{\theta_i}{2}\Big)}.
\end{equation}
It is easy to see that Eq.~\eqref{eq:eta2} is a special case of our result when all $\theta_i$ are equal $\frac{\pi}{N}$.

To simplify Eq.~\eqref{eq:eta4}, we set $\theta_N=\theta$. Then from Eq.~\eqref{eq:cond}, we have $\theta_i = \frac{\pi-\theta}{N-1}$ for $i\in[1,N-1]$. Following the result in Fig. 2c, the circuit is executed on the Valencia quantum computer provided by IBMQ \cite{IBMQ,github}. In Fig.~\ref{fig3} (c) we show theoretical curves from Eq.~\eqref{eq:eta4} in solid gray, and simulations of the quantum circuit in Fig.~\ref{fig3} (b) using a virtual machine as red dots. Experiment data from the Valencia device is shown as circular dots in different colors for $N = 2,3,4,5$, with device errors expected at 7.2\%, 10.3\%, 13.8\%, 19.0\%, respectively. The code is executed 10 times, each with 8192 shots, and error bars are calculated based on these deviations. The measurement data fit pretty well to theory for a small number of BSs, at $N=2$ and $3$. However, running the same quantum circuit on other systems, such as London or Ourense, yields substantial errors, even with $N = 2$ (data not shown). At larger $N =$ 4 or 5, the experiment requires entanglement of more qubits that demands a "deeper" circuit and more gates. Our data are quite off compared to theoretical and simulation results. This is not a surprising result, as the IBM's NISQ computers are known for noises and mostly limit to shallow circuits \cite{ShorIBM,GroverIBM,Gambetta17,Sierra5,IBMBell,IBMduality,IBMcat,DevittPRA}. Finding the origin for these errors is outside the scope of this work, as it requires in-depth information from the chip design, fabrication process, microwave setup, cryostat, and their inter-connections. Although being unreliable with quite many errors, it is impressive that these computers reach $\eta> 0.64$ for $N=3$ with $\theta$ around 0.6 $\pi$, and thus satisfactorily demonstrate the interaction-free measurement. At the classical limit, we expect $\eta= 0$, for such an effect should be impossible.

\section*{The Hardy's paradox}
\begin{figure}[h!]
\begin{center}
\includegraphics[width=3.4in,keepaspectratio]{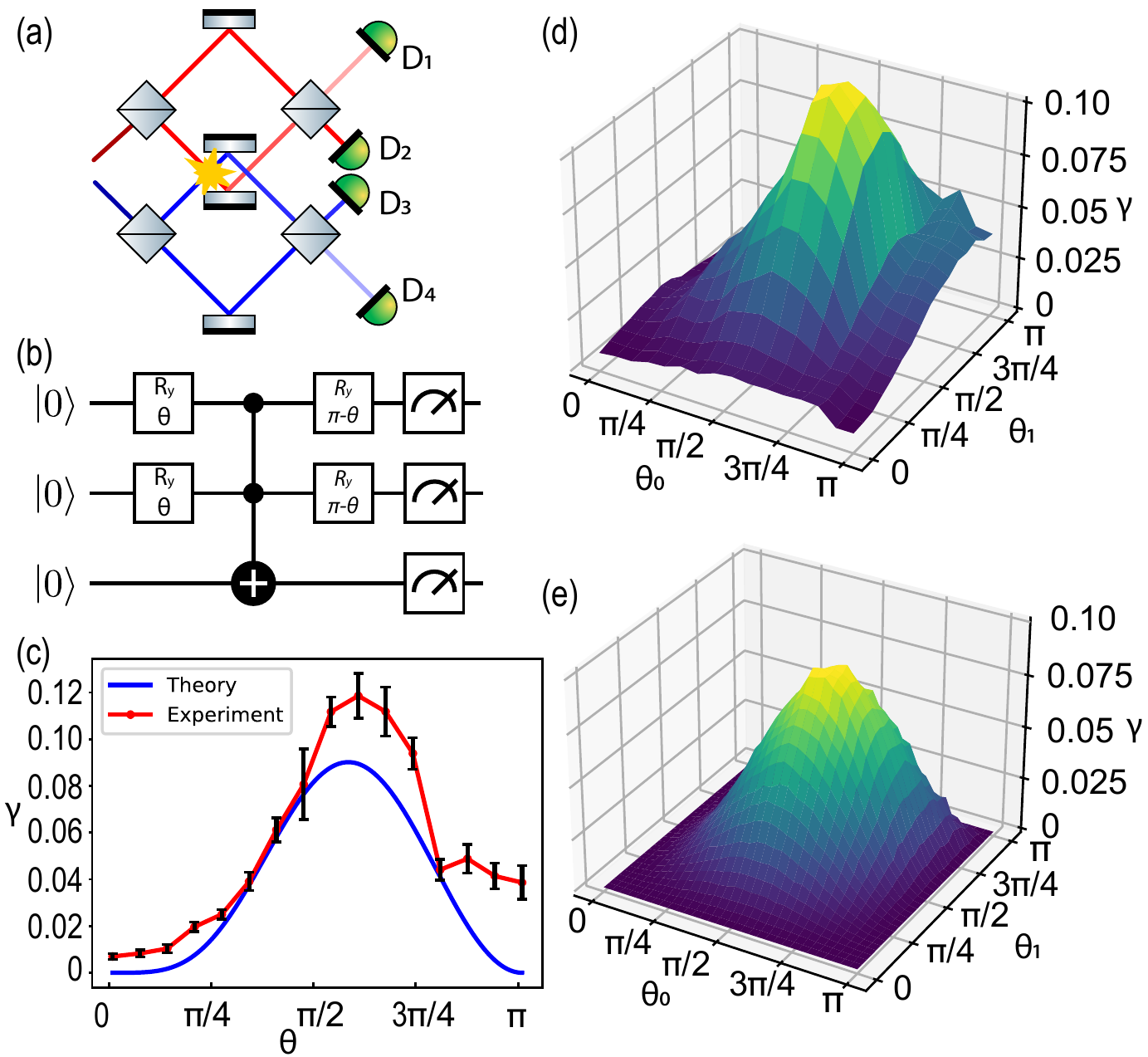}
     \caption{The Hardy's paradox experiment. (a) The original optical setup \cite{Hardy68} where two MZIs are combined. (b) The quantum circuit, where each particle passing through its BS is represented by a qubit evolves under the action of a $R_y(\theta_i)$ and a $R_y(\pi-\theta_i)$ gate. The annihilation cases are marked by the Toffoli gate. (c) Data from the IBM Vigo in direct comparison with theory when $\theta=\theta_0=\theta_1$. (d) The nonlocal probability as calculated from Eq.~\eqref{eq:gam6} showed in percentage. (e) Experiment data for the nonlocal probability obtained from the IBMQ Vigo quantum computer with a maximum of $\gamma=0.119$ at $\theta=0.533\pi$, close to Hardy's result \cite{Hardy71}. The relative device error is estimated at 12.1\% giving an absolute device error of $0.014$, while the largest standard deviation is 0.015.}
  \label{fig4}
\end{center}
\end{figure}

The original Elitzur-Vaidman bomb can be extended to illustrate the Hardy's paradox \cite{Hardy68,Hardy71} with the optical setup showed in Fig.~\ref{fig4} (a). Here, the classical bomb is replaced with two quantum objects, an electron and a positron. They enter two MZIs arranged so that their path crosses, which would lead to an annihilation typically. Without the positron, the electron would always go to its constructive interference detector D$_1$, and similarly for the case of the positron that light up detector D$_4$. A detection at the destructive interference detectors D$_2$ or D$_3$ indicates that a counterpart is sent through the MZI. The interesting case occurs when both D$_2$ and D$_3$ click, which indicates that the electron and the positron collide at the crossing point. According to their nature, they must annihilate and can not trigger D$_{2,3}$ detectors. The coincident clicks are a paradox that a local hidden-variable theory cannot explain \cite{Hardy68}. Nevertheless, the Hardy's paradox has been theoretically explained in terms of weak values \cite{Aha301,Kedem,Ho} and experimentally verified \cite{Irvine,Lundeen,ImotoNJP}. Hardy's paradox indicates the nonlocality feature of the quantum state inside the two MZIs \cite{RMPnonlocal}. Nonlocality is an intriguing topic that has been realized in photons \cite{Roger82,ZeilingerHardy,WhitePRL99,Luo18}, atoms \cite{Monroe,Weinfurter12}, or superconducting circuits \cite{Needley10,DiCarlo10}.

In Fig.~\ref{fig4} (b), the quantum circuit for this experiment is built similarly to the two previous experiments. Two qubits represent the two MZIs with rotation gates also satisfying Eq. (3) to put the qubits into their superpositions. In between the rotations, a Toffoli gate is targeted on the third qubit, revealing the quantum state of the control qubits when they are both at $|1\rangle$. Finally, the qubits are rotated back to the original basis for the measurement. With the original state denoted as $|\psi\rangle=|000\rangle$, the final state after the quantum circuit is
\begin{align*}
    |\psi_f\rangle=-&\frac14\sin\theta_1\sin\theta_0|000\rangle\\
    +&\frac12\sin\theta_1\sin^2\frac{\theta_0}{2}|100\rangle+\frac12\sin^2\frac{\theta_1}{2}\sin\theta_0|010\rangle\\
    +&\frac{1}{4}(2\cos\theta_1\sin^2\frac{\theta_0}{2}+\cos\theta_0+3)|110\rangle\\
    +&\frac14\sin\theta_1\sin{\theta_0}|001\rangle-\frac12\sin^2\frac{\theta_0}{2}\sin{\theta_1}|101\rangle\\
    -&\frac12\sin^2{\frac{\theta_1}{2}}\sin{\theta_0}|011\rangle+\sin^2\frac{\theta_1}{2}\sin^2{\frac{\theta_0}{2}}|111\rangle.\numberthis
\end{align*}
Here, $\theta_i$ is the rotation angle of the $i^{th}$ qubit, $i=[0,1]$. The state of interest, called the nonlocal state, corresponds to the case when the electron and the positron enter the cross-link without an annihilation. In our setup using a $R_y(
\theta_i)$ and a $R_y(\pi-\theta_i)$, it is state $|000\rangle$ where the first two-qubit has value $|q_0=0,q_1=0\rangle$ and the last qubit is 0: $|q_2=0\rangle$. To resemble the electron-positron annihilation in the quantum circuit,we reject all states with $|q_2=1\rangle$ in Eq. (5) and obtain the remain probabilities from  which the nonlocal probability for $|000\rangle$ is
\begin{align}\notag
    \gamma=&\frac{P(|000\rangle)}{P(|000\rangle)+P(|100\rangle)+P(|010\rangle)+P(|110\rangle)}\\
  =&\frac{\sin^2\theta_1\sin^2\theta_0}{4(2\cos\theta_1\sin^2\frac{\theta_0}{2}+\cos\theta_0+3)}\label{eq:gam6}\\
  =&\frac{2\sin^4\frac{\theta}{2}\cos^2\frac{\theta}{2}}{3-\cos\theta}.\label{eq:gam7}
\end{align}

Equation \eqref{eq:gam7} is obtained from Eq.~\eqref{eq:gam6} by choosing $\theta_1=\theta_0=\theta$, i.e., the two MZIs are set up identically. To show that our result is equivalent to Hardy's original probability, we change to new variables: $\sin\frac{\theta}{2}=\frac{\sqrt{\alpha\beta}}{\sqrt{1-|\alpha\beta|}}$ and $\cos\frac{\theta}{2}=\frac{|\alpha|-|\beta|}{\sqrt{1-|\alpha\beta|}}$. Eq.~\eqref{eq:gam7} becomes
\begin{align*}
  \gamma=&-2\bigg[\frac{|\alpha\beta|(|\alpha|-|\beta|)}{1-|\alpha\beta|}\bigg]^2\frac{1}{(|\alpha|-|\beta|)^2+2|\alpha\beta|-3}\\
  =&\bigg[\frac{|\alpha\beta|(|\alpha|-|\beta|)}{1-|\alpha\beta|}\bigg]^2,\numberthis
\end{align*}
which is Eq. 20 in Ref.~\cite{Hardy71}. The two approaches: the original calculation \cite{Hardy71} and the quantum circuit,
are indeed equivalent, which yield a maximum value $\gamma=\frac12(5\sqrt5-11) \approx 0.090$ at $\theta = 0.575\pi$. In Fig.~\eqref{fig4} (c), data from IBMQ Vigo are directly compared to theoretical result Eq.~\eqref{eq:gam6}, which show similar shapes of quite large error bars. These results are fully plotted in 3D as a function of the two beamsplitters' angles in Fig.~\eqref{fig4} (d) for theory and (e) for experiment. The maximum nonlocality probability $\gamma$ measures at $0.119\pm0.015$ around $\theta_0 = \theta_1 = \theta = 0.533\pi$, slightly larger than the value 0.090 estimated by Eq. 8 at $\theta_0 = \theta_1 = \theta = 0.575\pi$. This is the best result obtained by executing the quantum circuit in Fig. 4b multiple time on various IBM's quantum devices. Once again, the IBM quantum computer reproduces the Hardy's paradox reasonably well with decent precision, especially with a Toffoli gate that demands complicated technical details. 

\section*{Discussion and Conclusion}

\begin{table*}
	\begin{center}
	\begin{adjustbox}{width=1.5\columnwidth}		
	\begin{tabular}{|c|c|c|c|c|c|c|}\hline 
		Name & date & $T_1$ & $T_2$ & CNOT & Readout   & Connectivity  \\
		& mm/yy & ($\mu$s) & ($\mu$s) & error (\%) & error (\%) & \\
		\hline\hline
		Burlington & 08/20 & 84.88 & 67.36 & 1.50 & 4.64 & T   \\
		Essex & 08/20 & 104.31 & 123.7 & 1.76 & 3.59 & T   \\
		London & 08/20 & 61.45 & 62.74 & 1.75 & 4.40 & T   \\
		Ourense & 08/20 & 93.15 & 66.43 & 0.92 & 2.96 & T   \\
		Valencia (Fig. 2) & 08/20 & 84.18 & 62.78 & 1.11 & 2.32 & T   \\
		Valencia (Fig. 3) & 09/20 & 100.00 & 80.49 & 1.10 & 2.52 & T   \\
		Vigo (Fig. 1, 2) & 08/20 & 73.28 & 50.73 & 1.07 & 1.66 & T   \\
		Vigo (Fig. 4) & 09/20 & 107.64 & 74.04 & 0.94 & 1.96 & T   \\
		x2 & 08/20 & 57.08 & 45.40 & 1.82 & 3.18 & $\hourglass$  \\
		\hline
	\end{tabular}
	\end{adjustbox}
	\end{center}
	\caption{IBM quantum devices used in our experiments. "Date" indicates the time that we execute our codes, which can be found on Github \cite{github}. All other column including the lifetime $T_1$, decoherence time $T_2$, gate errors, readout errors are recorded at the time of the experiments. Their values are averaged over all 5 qubits. Connectivity denote the layout of the chip. Most of them are designed as T shape, which mean the center qubit connects to three other qubits.}
	\label{tab:calibration}
\end{table*}
The three chosen experiments all stemmed from the seeming non-definiteness of quantum observables \cite{EVBomb,Aha301,Bell,Kochen}. Inside an MZI where the photon's route is uncertain, it is the premise for restoration of the interference patterns in the eraser experiment, for the detection without interaction with the bomb, and for the collision at the cross-link of the electron-positron pair without an annihilation. As shown in Fig.~\ref{fig1} (a), Fig.~\ref{fig2} (a), Fig.~\ref{fig4} (a), these MZI apparatus are quite similar, but a slight variation in the experiment setup reveals profound quantum phenomena. Using an equivalent language, the quantum circuits showed in Fig.~\ref{fig1} (b), Fig.~\ref{fig2} (b), Fig.~\ref{fig4} (b) are also quite similar. Indeed, Fig.~\ref{fig1} (b) and Fig.~\ref{fig2} (b) are identical, implying a connection between the quantum eraser experiment and the Elitzur-Vaidman bomb. To see the similarity between the optical setup and a superconducting quantum processor, we should think of photon travels inside the MZI as a function of time. In the quantum circuit model, the quantum state evolves as a function of time under a series of gate actions, as we read it from left to right. In the original optical setup, these experiments explore the nonlocality of the photon path. In the superconducting setup, there is no spatial separation between the two eigenstates. Instead, the variable here is the charge number for charge qubits.

In 2020, our experiments are performed on some of IBM's quantum backends, namely Athens, Burlington, Essex, London, Ourense, Valencia, Vigo, and x2. These 5 qubit processors are of standard transmon design \cite{transmon,IBMQ} with lifetime T1 and decoherence time T2 are in the range of 50-150 $\mu s$, see Table 2 for more detail. They are equipped with an universal set of gates with fidelity in the range of 99\%. In their most popular layout, the center qubit in the T-shape processors like Vigo or Valencia has the most connectivity: it is in direct connection to three other ones. Although our quantum circuits are compiled from standard gates such as H, CNOT, or $R_y$, they are translated into the devices' physical gates denoted U1, U2, and U3. As noise arise from any possible sources, performance of these devices complicatedly depends on the chip design, fabrication process, cryostat setup, which are not available to end users. An in-depth analysis of these error is outside the scope of our work, but we do employ certain techniques to manage error and  achieve reasonable results. Firstly, the fidelity of a circuit is estimated as the product of fidelities of appliedgates given in that run, and the error of an experiment can be deduced thereafter. Secondly, the quality are tightly controlled as performances of these devices can fluctuate. Although calibrated daily, the data can suddenly be bad. Thirdly, the best qubits of the best machine are hand-picked for our circuit. Not all qubits are equal, and their qualities vary as IBM constantly updates their system. Finally, we applied mitigation error correction \cite{mitigation} extensively. Many times, this error-correction scheme can vastly improve our results. In any case, IBM's quantum computers are state of the art devices that offer an unprecedented opportunity to study the quantum world.

A central issue with gedanken quantum experiments is that they are difficult to verify on physical systems. The three experiments in our work have been realized on optical or atomic setups, the popular testbeds for foundation quantum experiments. However, we are unaware of any work that reproduces them on superconducting circuits, where the quantum object is the macroscopic collective state of condensates electron pairs. Replicating these experiments on the IBM quantum computers identifies two important aspects. First, it benchmarks the quantum nature of the IBM machines. Our results show that all experiments are reproduced with decent precision. The quantum eraser achieves pretty high accuracy, while the original bomb experiments yield results that fluctuate with devices, with the best machines being IBMQ Vigo and IBMQ Valencia. However, the accuracy is lost when the quantum circuit contains more than 4 qubits. Using the same codes and quantum circuit, we could not reproduce the Elitzur-Vaidman bomb in the general case when the qubit number is larger than 3. Apparently, we need gates with high fidelity, especially the 2 qubit gates that produce entanglement. Second, the IBM machines provide a general-purpose testbed for quantum experiments. Remotely controlled using a high-level language, we can reproduce a range of core quantum experiments in one run. By submitting a single job to the IBM quantum cloud, we execute both the quantum eraser, the Elitzur-Vaidman bomb, and the Hardy's paradox at once on a single solid-state device. More specially, we tune a wide range of parameters and thus verify our calculations as shown in the general bomb experiment in Fig.~\ref{fig3} (c). Similarly, in the Hardy's paradox experiment, our result spans the full range of $\theta_0, \theta_1$, which require a tremendous amount of work in the optical setup. Previously, such experiments needed weeks for the demonstration. As shown in our calculations above, we emphasize the equivalent between the quantum circuit language and the traditional analysis.

In summary, we present quantum circuits for the quantum eraser, the Elitzur-Vaidman bomb, and the Hardy's paradox experiments and prove that they are equivalent to the original setups using optical apparatuses. We further execute them on the IBM quantum computers and obtain reasonable results. While the current NISQ devices allow for the implementation of these experiments with few qubits, it is reasonable to demonstrate these proof-of-principle experiments as the testbed for testing quantum mechanics using noisy quantum computers.

This work was supported by JSPS, Japan KAKENHI Grant Number 20F20021.

\end{document}